\documentclass{icrc29}

\usepackage{graphicx,amssymb,amsmath,times}
\begin{document}

\title[Acceptance of fluorescence detectors for photons (...)]{Acceptance of fluorescence detectors for photons and its
implication in energy spectrum inference at the highest energies}

\author[Vitor de Souza, Gustavo Medina-Tanco and Jeferson A. Ortiz]
       {Vitor de Souza, Gustavo Medina-Tanco and Jeferson A. Ortiz \\ \\
{\it  Instituto de Astronomia, Geof\'{\i}sica e Ci\^encias Atmosf\'ericas, Universidade de S\~ao Paulo, Brasil} \\
}

\presenter{Presenter: V. de Souza (vitor@astro.iag.usp.br) bra-desouza-V-abs3-he14-poster}

\maketitle

\begin{abstract}
In this article, we investigate the acceptance of fluorescence
telescopes to different primary particles at the highest
energies. Using CORSIKA shower simulations without and with the new
pre-showering scheme, which allows photons to interact in the Earth
magnetic field, we estimate the aperture of the HiRes-I telescope for
gammas and protons primaries. We calculate the dependence of the
telescope sensitivity to primary particle identity. We also
investigate the possibility that systematic differences in shower
development for hadrons and gammas could mask or distort vital
features of the cosmic ray energy spectrum at energies above the
photo-pion production threshold. The impact of these effects on the
true acceptance of a fluorescence detector is analyzed in the context
of top-down production models.
\end{abstract}
%
%
\section{Introduction}\label{sec:introduction}

Top-down acceleration models, under very general conditions,
should produce a sizable signature of primary gammas. The
theoretical spectrum for such mechanisms shows an increase in the
flux for energies above 10$^{19.5}$~eV which could smooth away the
GZK cut-off of the hadronic component. HiRes \cite{bib:HiRes} and
AGASA \cite{bib:Agasa} experiments have explored the fluorescence
and the ground array experimental techniques, respectively, to
measure extensive air showers. Although AGASA shut down operations
in 2004, its results are still essential to investigate the
ultra-high energy cosmic rays. AGASA and HiRes have reported
considerable differences regarding the primary cosmic rays
spectrum around the GZK energies.

The ground array efficiency is based on a straightforward
detection given by its operational area which is, in principle,
not dependent on the primary particle type. On the other hand,
fluorescence telescopes measure the longitudinal air shower
development by detecting the fluorescence light emitted along the
track of particles in the atmosphere. Particles with different
mass induce showers with distinct longitudinal evolution in the
atmosphere and for this reason the efficiency of fluorescence
telescopes depends on the primary shower composition.

Since the original proposal of the Gaisser-Hillas function
\cite{bib:gaisser:hillas}, many studies
\cite{bib:hires:espectro,bib:song} have shown that the number of charged
particles (\emph{N}) in hadron-initiated showers as a function of
the atmospheric depth (\emph{X}) is well described by a four
parameter ($N_{\mathrm{max}}$, $X_0$, $X_{\mathrm{max}}$ and
$\lambda$) function. Recently, a physical process has been shown
to play an important role in the development of gamma-induced
showers. Primary gammas with energies above $10^{19}$~eV can be
converted into an electron-positron pair in the geomagnetic field
before entering the Earth atmosphere. The resultant
electron-positron pairs can lose their energy by emitting photons
due to magnetic bremsstrahlung. If the energy of the subsequently
emitted photon is high enough, another electron-positron pair can
be created. Hence an electromagnetic cascade (pre-shower) is
originated and will reach the Earth atmosphere instead of the
primary high-energy primary photon. The pre-shower effect is
significant for the development of the gamma shower in the
atmosphere, substantially modifying its longitudinal development
and making it look more similar to a nucleon-induced shower.

\begin{figure}[t]
\begin{minipage}[t]{7.5cm}
\includegraphics*[angle=-90, width=1.0\textwidth]{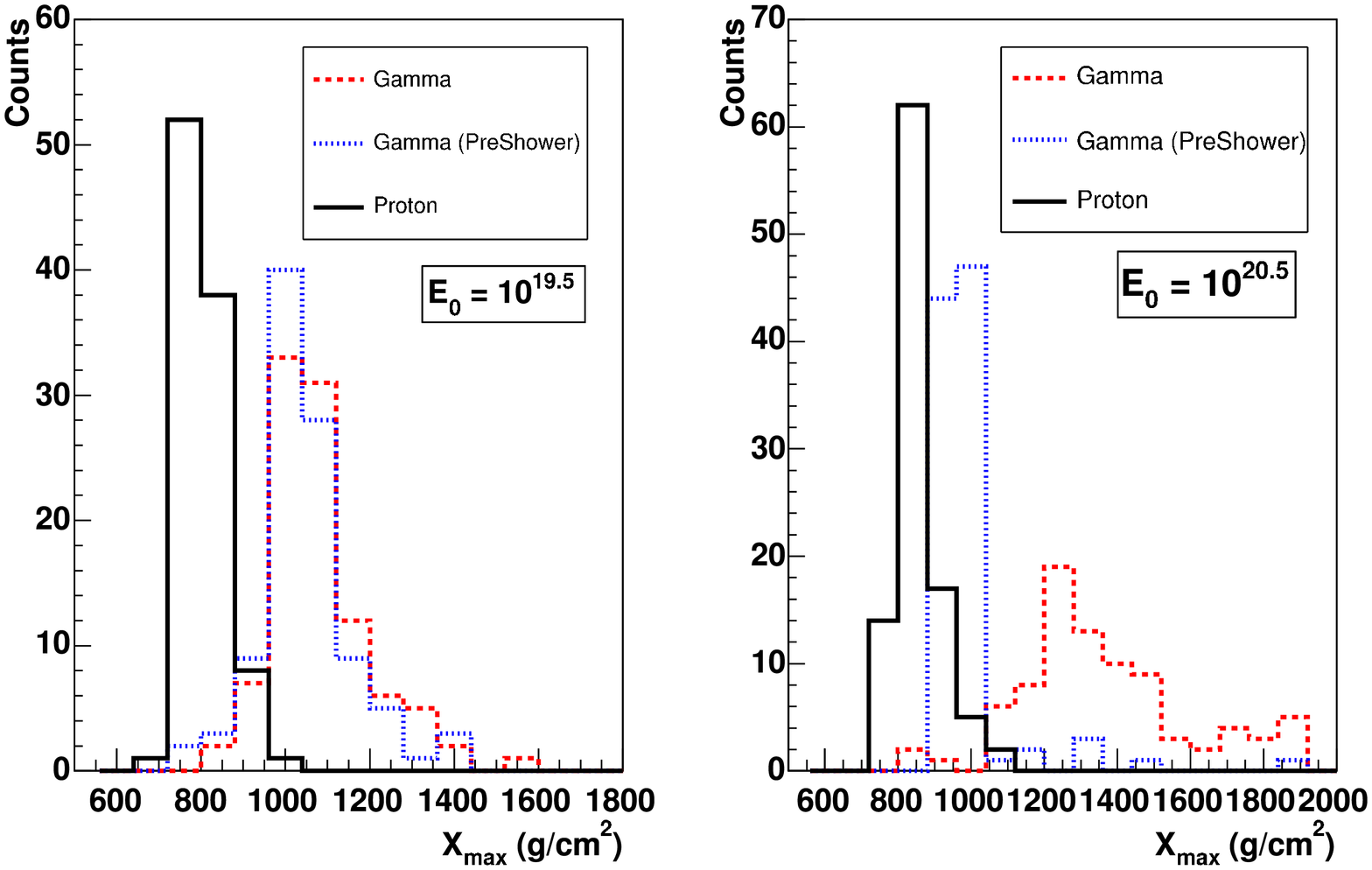}
\caption{\label{fig:xmax} Distribution of $X_{\mathrm{max}}$ for showers
   initiated by 
   protons and gammas with and without
   pre-shower. Results are shown for 100 showers at $45^\circ$ zenith angle
   and energies of    $10^{19.5}$ and $10^{20.5}$~eV.}  
\end{minipage}
\hfill
\begin{minipage}[t]{7.5cm}
\includegraphics*[angle=-90,width=1.0\textwidth]{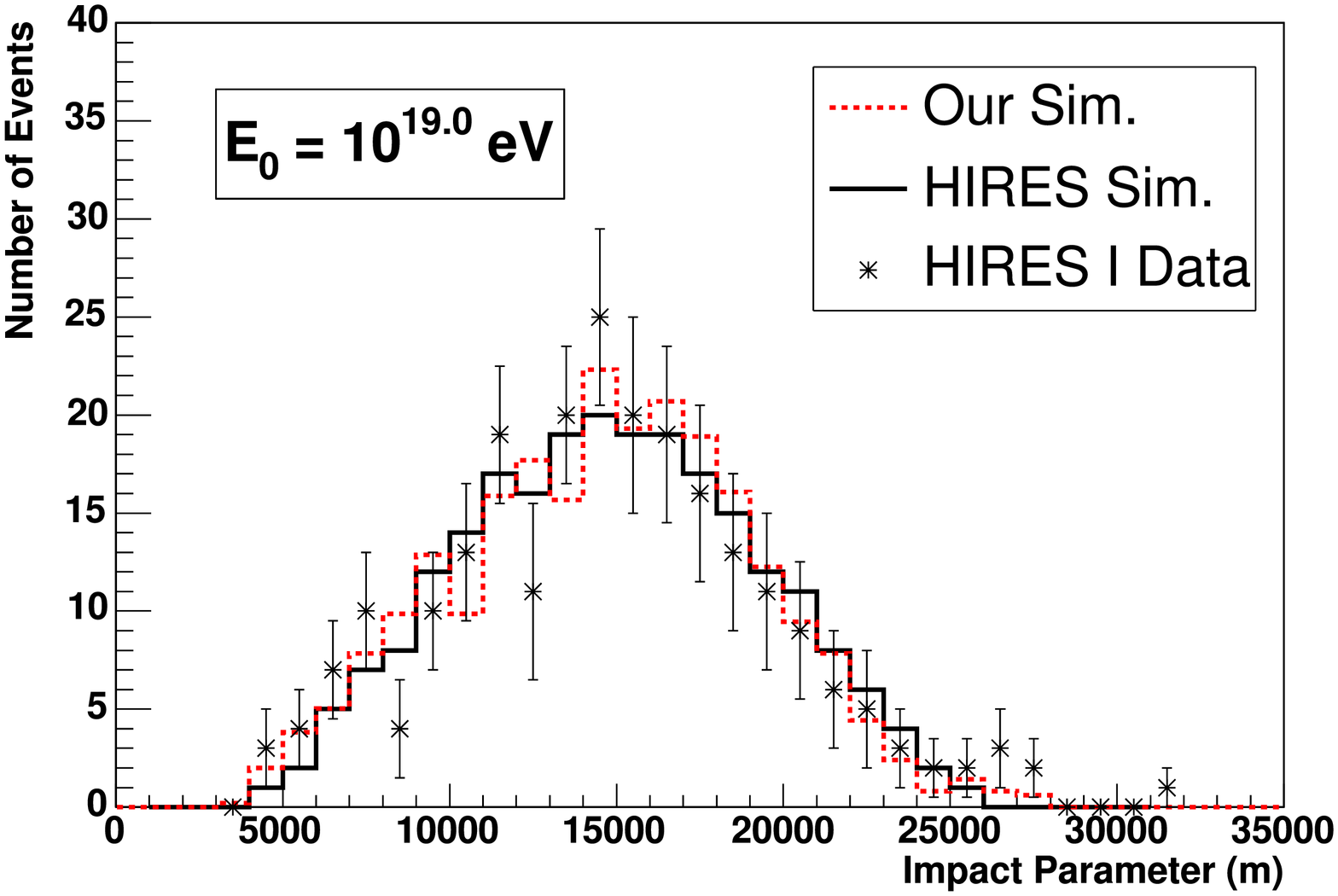}
\caption{\label{fig:rp} Distribution of impact parameters for showers
   at primary energies of $10^{19}$~eV. We compare our simulation with
   HiRes-I data and simulation}  
\end{minipage}
\end{figure}


Fig.~\ref{fig:xmax} shows the $X_{\mathrm{max}}$ distribution of
100 proton and gamma initiated showers. The influence of the
pre-shower effect on gamma showers is illustrated by the
$X_{\mathrm{max}}$ distribution at $10^{19.5}$ (left panel) and
$10^{20.5}$~eV (right panel). At $10^{19.5}$~eV the
$X_{\mathrm{max}}$ distributions for gamma showers with (dotted
line) and without (dashed line) the pre-shower are almost
identical and we can verify a not significant reduction of the
average $X_{\mathrm{max}}$ distribution value. However, at
$10^{20.5}$~eV the pre-shower influence is easily perceived,
producing a quite reasonable difference between the
$X_{\mathrm{max}}$ distributions obtained by simulations with and
without the pre-shower effect.
%
%
\vspace{-0.8cm}
\section{Shower and Detector Simulation}\label{sec:simulation}

In this contribution we used Monte Carlo simulations to evaluate
the HiRes-I telescope aperture, following in detail the general
specifications published by the HiRes collaboration in
\cite{bib:hires:espectro}.

According to the HiRes collaboration procedures, we have performed
a very detailed simulation of the cosmic ray air shower and the
HiRes-I telescope. Proton and gamma ray showers have been
simulated with the well tested CORSIKA \cite{bib:corsika} code,
using the QGSJet hadronic interaction model \cite{bib:qgsjet}, for
the energy range $10^{19}$-$10^{20.5}$~eV, in steps of $0.1$~dex.
For each energy and primary particle we have generated 100 events.
A recent release of CORSIKA (version 6.2) implements the
pre-shower effect \cite{bib:pre:shower}. For gamma ray showers, we
have simulated two sets: a) with the pre-shower effect b) without
the pre-shower effect. In order to save computation time, each
CORSIKA simulated shower has been recycled several times by
randomly drawing zenith angles and core positions.

The HiRes-I telescope has been simulated with a specific
simulation program which calculates the number of fluorescence
photons along the shower path, from the longitudinal development
of the charged particles simulated by CORSIKA, and propagates the
photons through the telescope. Fig. \ref{fig:rp} shows a
comparison between our simulation program and the HiRes-I data and
the HiRes collaboration simulation. It is possible to verify the
good agreement between our simulation and the HiRes data, showing
that we are able to reproduce the most important features of the
detector. The HiRes data and simulation have been extracted from
\cite{bib:hires:espectro}. We have calculated the impact parameter
of the showers detected by the telescopes. The number of simulated
showers was normalized to the number of detected events.

\begin{figure}[t]
\begin{minipage}[t]{7.5cm}
\includegraphics*[angle=-90,width=1.0\textwidth]{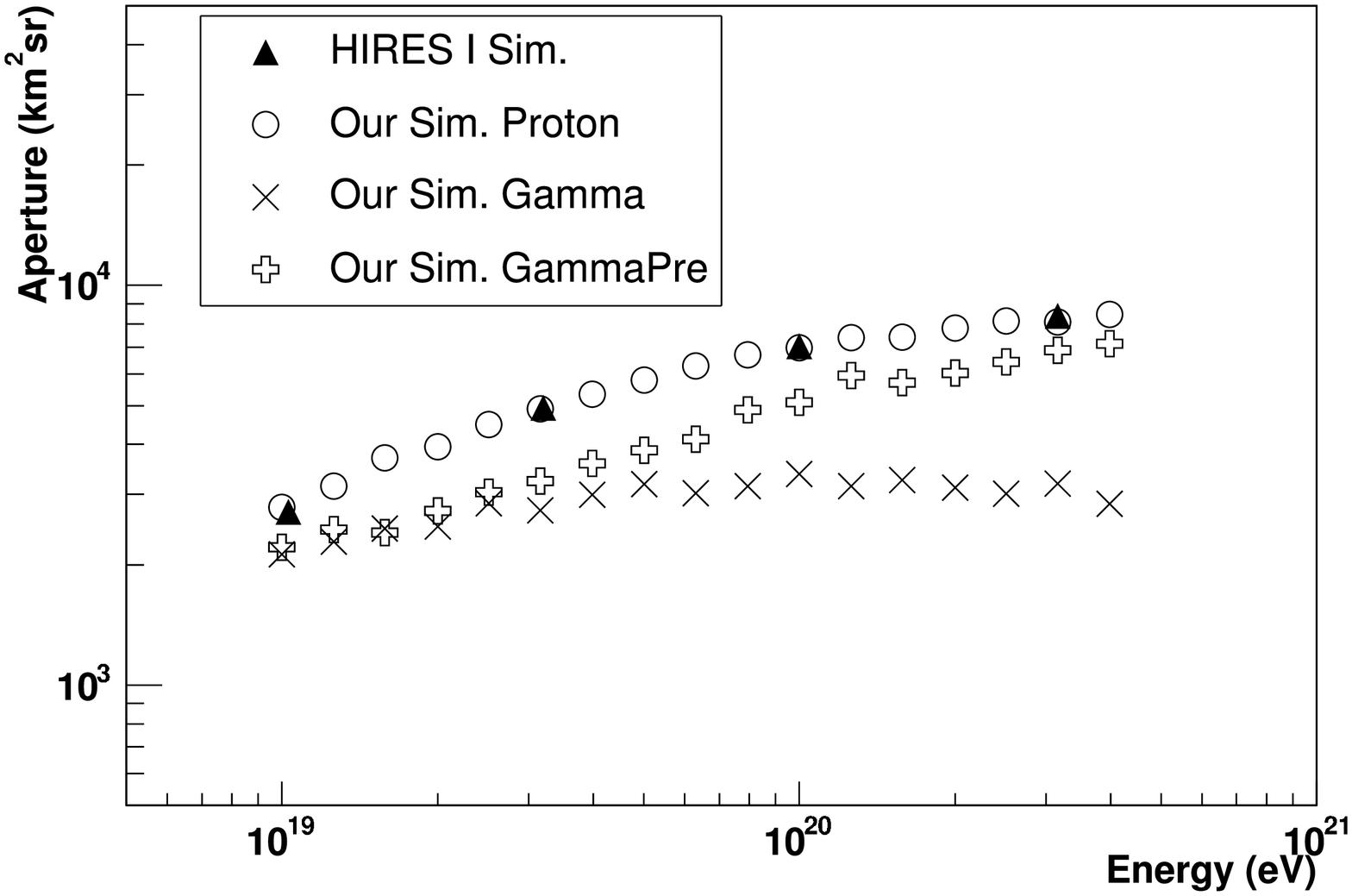}
\caption{\label{fig:aperture} HiRes-I telescope aperture calculated
   by the HiRes-I collaboration and published in reference
   \cite{bib:hires:espectro}, calculated with our analysis program
   using shower generated by the CORSIKA code.}
\end{minipage}
\hfill
\begin{minipage}[t]{7.5cm}
\vspace{0.6cm}
\includegraphics*[trim= 4cm 5cm 4cm 4cm,angle=-90,width=1.0\textwidth]{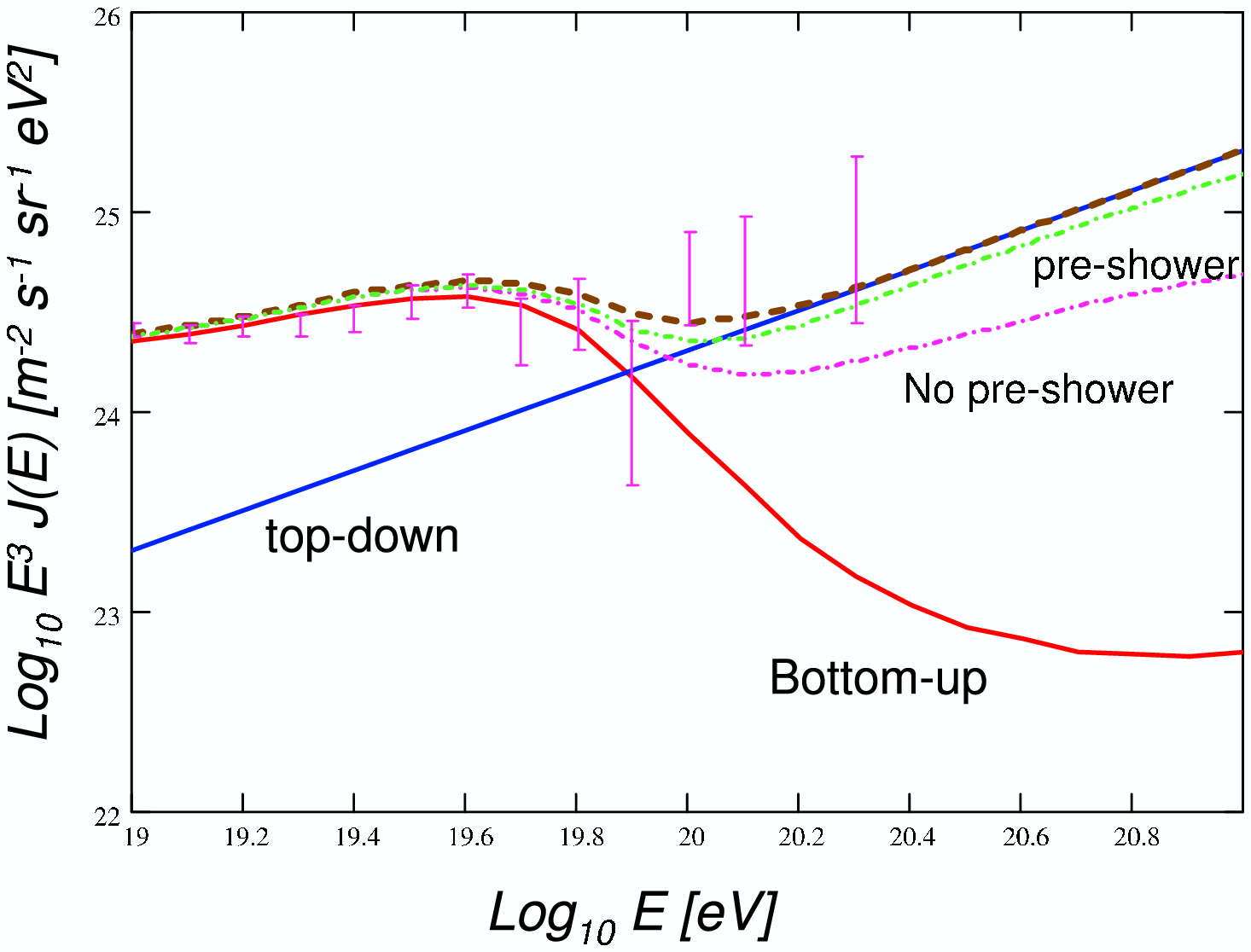}
\caption{\label{fig:espectro} Cosmic ray spectrum as seen by the
  HiRes-I telescope for different primary particles and production scenarios.}
\end{minipage}
\end{figure}

%
%
\vspace{-0.5cm}
\section{Reconstruction and Analysis}\label{sec:reconstruction}

The reconstruction of the showers also followed in detail the
particular procedures described in references
\cite{bib:hires:espectro}. Basically, the
inverse process described in section \ref{sec:simulation} is
applied to reconstruct the shower longitudinal profile and
consequently its energy. The number of photoelectrons measured in
each pixel of the detector is mapped backwards onto the number of
photons in the telescope, the number of photons in the axis of the
shower and, finally, the number of particles in the shower. The
telescope efficiency, atmospheric absorption, fluorescence yield
and missing energies have been considered according to section
\ref{sec:simulation} and references there in.

A Gaisser-Hillas \cite{bib:gaisser:hillas} function was fitted to
the data in agreement to the specifications in reference
\cite{bib:hires:espectro}. The $X_{\mathrm{max}}$ parameter was
allowed to vary in 35~g/cm$^2$ steps between 680 and 900~g/cm$^2$.
The $X_0$ parameter of the Gaisser-Hillas was fixed to
-~60~g/cm$^2$, in conformity to reference
\cite{bib:hires:espectro:2}. Quality cuts are always needed to
ensure an accurate reconstruction. We have required events to
satisfy the conditions listed below as taken from reference
\cite{bib:hires:espectro}: 1) Average number of photoelectrons per
phototube greater than 25; 2) Angular speed less than $3.33^\circ \ \
\mu s$; 3) Track arc-length greater than $8.0^\circ$; 4) Depth of
first observed point less than 1000~g/cm$^2$; 5) Angle of the shower
in the plane containing the shower axis and the detector greater than
$120^\circ$. Showers which did not obey those conditions were rejected, and
excluded from further analysis.


%
%
\vspace{-0.5cm}
\section{HiRes-I Telescope Aperture and Spectrum}\label{sec:aperture}

We have investigated the dependence of the aperture on different
primary particles. According to \cite{bib:hires:espectro} the
HiRes collaboration has determined the spectrum based only on
aperture investigations of the HiRes-I telescope for proton and
iron showers.

Fig. \ref{fig:aperture} shows the aperture for proton and gamma
initiated showers (with and without the pre-shower effect). In all
curves, we have used 100 different showers shuffled 50 times each
as explained before in section \ref{sec:simulation}.


If the pre-shower effect is not taken into account, gamma showers
tend to develop much deeper in the atmosphere when compared to
proton induced showers. This fact makes gamma ray showers harder
to detect, which represents a great reduction in the telescope
aperture for all energies.

When the pre-shower effect is considered, gamma showers with energy
above $10^{19.5}$~eV have a considerable probability of conversion
into a electron-positron pair (more than 5\%)
\cite{bib:pre:shower}. The probability conversion increases very
rapidly with increasing energy and reaches 100\% between
$10^{20.0}$ and $10^{20.5}$~eV depending on the arrival direction
of the particles related to the Earth magnetic field. The aperture
calculation shown in fig.~\ref{fig:aperture} illustrates this
increase of the conversion probability. Gamma showers simulated
with the pre-shower effect evolve from a ``gamma without
pre-shower profile'' to a ``hadronic profile'' with energies
varying from $10^{19.5}$ to  $10^{20.6}$~eV, which make the
HiRes-I aperture for gamma ray showers close to aperture for the
hadronic showers.

Despite the fact that the conversion probability of a gamma in the
Earth magnetic field reaches 100\% at $10^{20.6}$~eV, the HiRes-I
telescope aperture for gamma ray showers is smaller than the
aperture for proton showers.

\vspace{-0.5cm}
\section{Astrophysical significance}

The fact that the HiRes aperture has been calculated under the
assumption of hadronic primaries, opens the possibility of the
existence systematic effects for a broad range of cosmic ray
production scenarios. In particular, it cannot be disregarded at
present the possibility of mixed extragalactic components: a
hadronic one, coming from conservative bottom-up acceleration
mechanisms and a photon, harder component originated in more
exotic top-down models.

As an example, figure ~\ref{fig:espectro} illustrates such a
combination of spectra. The GZK-ed spectrum (lower thick line) has
been numerically calculated using a homogeneous distribution of
cosmological sources whose luminosity evolves with redshift
according to $(1+z)^m$ with $m=3$. Protons were injected at the
sources,with a power law spectrum (spectral index $\nu_{i}=2.7$
and energy loses due to redshift, pair creation and photo-pion
production in interactions with the cosmic microwave radiation are
included. The resultant spectrum is normalize to the AGASA
observed flux at $10^{19}$ eV. The normalization of the harder
(spectral index $\nu_{TD}=2$) top-down spectrum is such that
trans-GZK AGASA data can be fitted by the combined spectrum (thick
dashed line).

The dot dashed curves illustrate the spectrum that HiRes would
infer for the case of a photon component described with and
without pre-showering. It can be seen that in the case without
pre-shower the effect could be severe enough as to make AGASA and
HiRes spectra compatible within quoted uncertainties. The
existence of pre-showers considerably diminishes this effect,
maintaining unaltered the AGASA-HiRes discrepancy.

%
\vspace{-0.5cm}
\section{Acknowledgments}

This paper was partially supported by the Brazilian Agencies CNPq
and FAPESP. Most simulations were carried out on a Cluster Linux
TDI, supported by Laborat\'orio de Computa\c c\~ao Cient\'{\i}fica
Avan\c cada at Universidade de S\~ao Paulo.

%
%

\end{document}